# Effect of biaxial strain and hydrostatic pressure on the magnetic properties of bilayer CrI$_3$

Chong Xu[1], Qianjun Wang[1], Bin Xu[1], Jun Hu[1,2,*]

[1]School of Physical Science and Technology & Jiangsu Key Laboratory of Thin Films, Soochow University, Suzhou 215006, China.

[2]School of Physical Science and Technology, Ningbo University, Ningbo 315211, China.

Two-dimensional van der Waals magnetic materials are intriguing for applications in the future spintronics devices, so it is crucial to explore strategy to control the magnetic properties. Here, we carried out first-principles calculations and Monte Carlo simulations to investigate the effect of biaxial strain and hydrostatic pressure on the magnetic properties of the bilayer CrI$_3$. We found that the magnetic anisotropy, intralayer and interlayer exchange interactions, and Curie temperature can be tuned by biaxial strain and hydrostatic pressure. Large compressive biaxial strain may induce a ferromagnetic-to-antiferromagnetic transition of both CrI$_3$ layers. The hydrostatic pressure could enhance the intralayer exchange interaction significantly and hence largely boost the Curie temperature. The effect of the biaxial strain and hydrostatic pressure revealed in the bilayer CrI$_3$ may be generalized to other two-dimensional magnetic materials.

---

[*] Corresponding author. Email: hujun2@nbu.edu.cn

## I. Introduction

Van der Waals two-dimensional (2D) materials have evoked great interest of researchers in the community of condensed matter physics and materials science, since the discovery of graphene [1], because they exhibit abundant new physical properties, and are promising in various applications in electronic and optoelectronic devices [2]. Recently, a new class of 2D materials—2D ferromagnetic (FM) materials—have been discovered [3,4]. The 2D FM materials are particularly interesting not only in the aspect of fundamental science, because it was unexpected according to the Mermin-Wagner theorem [5], but also in the development of new generation of spintronics devices, because the highest Curie temperature ($T_C$) may even achieve to room temperature [6,7,8,9].

The bulk $CrI_3$ is FM with the $T_C$ of 61 K and a rhombohedral layer stacking [10,11]. The FM ordering persists in few-layer $CrI_3$ except bilayer, with the $T_C$ down to 45 K for the monolayer $CrI_3$, whereas the bilayer $CrI_3$ displays interlayer antiferromagnetic (AFM) coupling [4]. It is known that the physical properties may be closely associated with the stacking order in 2D matertials [12]. Indeed, a series of literatures have reported that the interlayer magnetic coupling in the bilayer $CrI_3$ depends on the stacking pattern [13,14,15,16]. With the AB-stacking order (Fig. 1a) where the top layer laterally shifts by $[\frac{2}{3},\frac{1}{3}]$ in fractional coordinates with respect to the bottom layer, the two $CrI_3$ layers couple ferromagnetically. With the AB'-stacking order (Fig. 1b) where the lateral shift is $[\frac{1}{3},\frac{1}{3}]$, the two $CrI_3$ layers couple antiferromagnetically. In fact, these two types of stacking orders correspond to the bulk $CrI_3$ at low temperature (below 210 K) with the space group $R\bar{3}$ and room temperature with the space group $C2/m$, respectively [11,17]. Nevertheless, the AB'-stacking bilayer $CrI_3$ was usually observed in experiment, probably because the bilayer exfoliated at room temperature is kinetically trapped in the AB'-stacking order during the cooling process [13]. Interestingly, the interlayer magnetism of the bilayer $CrI_3$ can be tuned by electric field, gate voltage and external pressure [17,18,19,20,21].

Magnetic anisotropy (MA) is an important requirement for realizing long-rang magnetic ordering in 2D magnetic materials [4,22], because the long-rang 2D magnetic ordering cannot establish in the spin-rotational invariant systems as demonstrated in the Mermin-Wagner theorem [5]. Magnetic anisotropy plays the role of breaking the spin rotational invariance, and thus allows the long-rang 2D magnetic ordering to exist. For the sake of practical applications in spintronics devices, perpendicular magnetic anisotropy (PMA) magnetic and FM ordering at room temperature are required simultaneously. Therefore, strong PMA and FM ordering are desired and it is important to explore strategies to control the MA and to enhance $T_C$ [23]. Although theoretical study of the effect of strain on the MA of the monolayer $CrI_3$ has been done recently [24,25,26,27], systematic theoretical investigation of the electronic and magnetic properties including the MA and magnetic ordering of the bilayer CrI$_3$ under strain is still lacking [28].

In this paper, we explored the effect of biaxial strain and hydrostatic pressure on the electronic and magnetic properties of the bilayer CrI$_3$, based on first-principles calculations and Monto Carlo simulation. We found that the magnetic properties of the bilayer CrI$_3$ are sensitive to biaxial strain and hydrostatic pressure. In particular, the hydrostatic pressure may turn the easy axis from the out-of-plane direction (i.e. PMA) to in-plane direction. The intralayer exchange interaction may be significantly enhanced by the hydrostatic pressure, which may lead to T$_C$ over 100 K, much higher than the original T$_C$ without the hydrostatic pressure.

## II. Computational details

The atomic structures, electronic and magnetic properties were calculated by using the Vienna *ab-initio* simulation package in the framework of the density functional theory (DFT) [29,30]. The interaction between valence electrons and ionic cores was described within the framework of the projector augmented wave (PAW) method [31,32]. Van der Waals interaction was included with the optB86b exchange functional [33,34]. Hubbard U of 3 eV for the Cr atom was adopted to account for the strong electronic correlations [35]. The energy cutoff for the plane wave basis expansion was set to 650 eV. The two-dimensional Brillouin zone was sampled by a 24×24 k-grid mesh. The atomic positions were fully relaxed using the conjugated gradient method until the force on each atom is smaller than 0.01 eV/Å.

The exchange coupling parameters of the bilayer CrI$_3$ can be extracted by comparing the total energies with different spin configurations (see Fig. 2), based on the Heisenberg spin Hamiltonian

$$H = -\sum_{i,j} J_{ij} \vec{S}_i \cdot \vec{S}_j. \tag{1}$$

As illustrated in Fig. 1, we considered the first nearest neighboring intralayer exchange interaction (denoted by J$_1$) for both AB- and AB'-stacking bilayers. For the interlayer couplings, the dominant exchange interactions can be represented by J$_2$ and J$_3$ for the AB-stacking bilayer CrI$_3$ (Fig. 1a) and J$_2$ for the AB'-stacking bilayer CrI$_3$ (Fig. 1b). With these exchange interaction parameters, we performed Monte Carlo simulations with a 64×64 supercell to estimate the T$_C$.

## III. Results and discussion

We started our calculations from the experimental lattice constant of the bulk CrI$_3$ (a$_0$ = 6.867 Å), and then applied in-plane biaxial strain (ε) from -6% to 6%. The negative and positive signs indicate compressive and extensile strains, respectively. The corresponding lattice constant varies from 6.45 Å to 7.28 Å. Accordingly, the distance between the nearest Cr atoms increases from 3.73 Å to 4.20 Å, and the Cr-I bond lengths increase from 2.71 Å to 2.79 Å, as plotted in Fig. 3a. Clearly, the variation of the Cr-I bond lengths (from -1.45% and 1.45%) is much smaller than the in-plane biaxial strain, because the out-of-plane positions of the Cr and I atoms are free to relax. Consequently, the Cr-I-Cr angle ($\alpha$) increases significantly when the strain varies from -6% to 6%, as shown in Fig. 3b. There are two types of I-Cr-I angles, the angle formed by one Cr atom and two I atoms lying in the same atomic plane ( two atomic planes) is denoted as $\theta_1$ ($\theta_2$). $\theta_1$ increases but slower than $\alpha$, while $\theta_2$ decreases significantly, as seen in Fig. 3b. Note that these structural parameters do not depend on the stacking patterns.

The bilayer $CrI_3$ is semiconducting with the band gap of 1.01 and 1.07 eV, respectively for the AB- and AB'-stacking bilayers, slightly smaller than that of the monolayer $CrI_3$ (1.09 eV). Generally speaking, the band gap of a semiconductor will increase (decrease) under compressive (extensile) strain, as sketched in Fig. 4a. However, the band gap of the bilayer $CrI_3$ exhibits opposite trend. As shown in Fig. 4c and 4d, the band gaps of the AB- and AB'-stacking bilayers respectively decrease to 0.74 and 0.69 eV under compressive strain of -6%, while they increase to 1.13 and 1.22 eV under extensile strain of 6%. This abnormal phenomenon can be explained by the feature of the local structure which is represented by the distorted octahedron formed by one Cr atom and six I atoms as displayed in Fig. 4b. In the local coordinate frame, the 3d orbitals of the Cr atom split into the triplet $t_{2g}$ orbital and the doublet $e_g$ orbital due to the crystal field, as seen from the inset in Fig. 4c. The band gap lies between the occupied $t_{2g}$ states and the empty $e_g$ states, and thus is primarily determined by the crystal-field splitting. When a compressive (extensile) biaxial strain is applied onto the unit cell, the equivalent strain onto the octahedron is actually extensile (compressive) along the $C_3$ axis as indicated in Fig. 4b. Therefore, the compressive (extensile) biaxial strain leads to the octahedron to be stretched (compressed), and thus weakens (strengthens) the crystal-field splitting. Consequently, the energy levels of the $e_g$ states decrease (increase), resulting in the decrease (increase) of the band gap. Clearly, the evolution of the band gap under external strain still accords with the trend shown in Fig. 4a, if we consider the equivalent strain on the octahedron rather than the biaxial strain on the unit cell.

The biaxial strain also tunes the magnetic properties. As shown in Fig. 5, the local spin moment contributed by the Cr atoms ($M_{S,Cr}$) increases as the strain change from -6% to 6%. In other words, the larger lattice constant results in the larger $M_{S,Cr}$. The total increment of $M_{S,Cr}$ is about 0.1 $\mu_B$. However, the total spin moment of each $CrI_3$ formula is independent of the strain and maintains 3.0 $\mu_B$. It is clear that the value of $M_{S,Cr}$ is larger than the total spin moment, because the I atoms contribute about -0.1 $\mu_B$ per atom. Note that the stacking patterns do not affect the spin moments. The magnetic anisotropy energy (MAE) are positive for all strains, indicating PMA for the bilayer $CrI_3$, in agreement with experimental observations [4]. In addition, the MAE increases significantly when the amplitude of the compressive strain increases, while it decreases slightly when the extensile strain increases, similar with a recent report [26]. Furthermore, the MAE of the AB-stacking bilayer $CrI_3$ is larger than that of the AB'-stacking bilayer $CrI_3$ in most range of the strains. At the experimental lattice constant (i.e.

ε = 0), the MAEs of the AB- and AB'-stacking bilayers are 0.58 meV and 0.34 meV per $CrI_3$ formula, respectively, in agreement with the recent report [23]. When ε = -6%, the MAE becomes 1.4 meV per $CrI_3$ formula for both AB- and AB'-stacking bilayers. Obviously, larger MAE guarantees better thermal stability at high temperature.

Note that the properties discussed above do not depend on the interlayer exchange interaction. The two $CrI_3$ layers may couple with each other either ferromagnetically or antiferromagnetically [4]. The ground state can be determined by calculating the energy difference between the FM and AFM states: $\Delta E = E_{AFM} - E_{FM}$. As shown in Fig. 6a, the AB-stacking bilayer $CrI_3$ manifests strong interlayer ferromagnetism for the whole range of the strain. On the contrary, a FM-to-AFM transition occurs near ε = -2% for the AB'-stacking bilayer $CrI_3$. With compressive strain ε ≤ -2%, the two $CrI_3$ layers couple with each other ferromagnetically with the magnetic configuration in Fig. 2a. The interlayer coupling becomes AFM for minor compressive strain and all extensile strain, and the corresponding magnetic configuration is similar with that plotted in Fig. 2b. Therefore, the ground state of the AB'-stacking bilayer $CrI_3$ is interlayer AFM, in agreement with the experimental observation [4].

To reveal the underlying mechanism of the exchange couplings under the biaxial strain, we calculated the total energies of four different magnetic configurations as plotted in Fig. 2. In the framework of Heisenberg model given by Eq. (1), the total energies related to the exchange couplings for the four different magnetic configurations in Fig. 2 for the AB-stacking bilayer $CrI_3$ can be written as follows: $E_a = -(6J_1 + J_2 + 9J_3)S^2$; $E_b = -(6J_1 - J_2 - 9J_3)S^2$; $E_c = -(J_2 - 3J_3)S^2$; $E_d = +(6J_1 + J_2 - 3J_3)S^2$; $E_e = +(6J_1 - J_2 + 3J_3)S^2$. For the AB'-stacking bilayer $CrI_3$, the total energies are $E'_a = -(6J_1 - 4J_2)S^2$; $E'_b = -(6J_1 + 4J_2)S^2$; $E'_c = 0$; $E'_d = +6J_1S^2$; $E'_e = +6J_1S^2$. Here, S = 3/2 is the spin quantum number of each $CrI_3$ formula. Then we can obtain these exchange interaction parameters.

Figure 6b plots the exchange interaction parameters. It can be seen that the intralayer exchange interaction energy ($J_1$) increases when the extensile strain increases and decreases when the compressive strain increases, for both AB- and AB'-stacking bilayers. The amplitudes of $J_1$ of the AB- and AB'-stacking bilayers are almost the same. In other words, the intralayer exchange interaction in bilayer $CrI_3$ depends on the lattice constant monotonically but is independent of the stacking patterns. The larger lattice constant results in the larger $J_1$. In addition, the sign of $J_1$ changes around ε = -6%, which indicates a intralayer FM-to-AFM transition. Similar magnetic phase transition was also observed in the monolayer $CrI_3$ [26,27]. With the extensile strain (ε > 0) and moderate compressive strain (-6% < ε < 0), the neighboring Cr atoms in the same $CrI_3$ layer couple with each other ferromagnetically. Combined the ferromagnetic and antiferromagnetic interlayer couplings respectively for the AB- and AB'-stacking bilayers, the AB- and AB'-stacking bilayers adopt the magnetic configurations displayed in Fig. 2a and 2b, respectively, in this situation. With large compressive strain (ε ≤

-6%), the intralayer exchange interaction becomes antiferromagnetic. Since the interlayer coupling is ferromagnetic for both AB- and AB'-stacking bilayers, they adopt the magnetic configuration shown in Fig. 2e. This magnetic configuration is the same as that reported recently [28]. The trend of the strain-dependent exchange interactions can be understood qualitatively as follows. It has been revealed that $J_1$ contains two parts of contributions: the direct exchange interaction between the first nearest neighboring Cr atom via $t_{2g}$-$t_{2g}$ orbitals and the indirect superexchange interaction via the $t_{2g}$-p-$e_g$ orbitals [23,27] Because of the octahedral crystal-field splitting, the former favors AFM exchange interaction, while the later favors FM exchange interaction. With extensive strain and moderate compressive strain, the indirect superexchange interaction dominates the intralayer exchange interaction in the $CrI_3$ layer, which results in the FM ground state and positive $J_1$. Furthermore, when the lattice constant increases, the Cr-Cr distance and the Cr-I-Cr angle ($\alpha$ in Fig. 3) increases. Consequently, the direct exchange interaction is weakened, while the superexchange interaction is enhanced due to the stronger $t_{2g}$-p hybridization. These combined effects make $J_1$ increases. With large compressive strain $\varepsilon \leq$ -6%), the Cr-Cr distance is short enough, so that the Cr-Cr direct exchange interaction dominates, resulting in AFM ground state and negative $J_1$. Moreover, when $\varepsilon$ = -6%, the MAEs of the AB- and AB'-stacking bilayers are -1.60 and -2.15 meV, respectively, as indicated by open symbols in Fig. 5, implying in-plane MA rather than PMA.

For the interlayer exchange interactions of the AB-stacking bilayer $CrI_3$, $J_2$ is negative and its amplitudes in the most range of the strain are large, indicating that the two Cr atoms connected by $J_2$ (see Fig. 1) prefer AFM coupling. $J_3$ is positive and its amplitude is smaller than that of $J_2$. For each Cr atom, there is only one pair of exchange interaction represented by $J_2$, while there are six pairs of exchange interaction represented by $J_3$. Therefore, the final interlayer exchange interaction is ferromagnetically. For the AB'-stacking bilayer $CrI_3$, the amplitudes of $J_2$ are relatively small, and the sign changes around $\varepsilon$ = -2%, corresponding to the FM-to-AFM transition as shown in Fig. 6a.

With the exchange interaction parameters, we carried out Monte Carlo (MC) simulation based on the Heisenberg model as expressed in Eq. (1). Although the bulk $CrI_3$ is believed to be Ising type FM magnet [4] we found that the Ising model overestimates $T_C$ greatly. For the monolayer $CrI_3$, the Curie temperatures from the Ising model, Heisenberg model, and mean-field theory are 130 K, 90 K and 52 K, respectively. Apparently, only the mean-field theory predicts proper $T_C$ close to the experiment measurement (45 K). Nevertheless, the mean-field theory could not be applied to the bilayer $CrI_3$ due to the complicated exchange interactions. Therefore, we choose the Heisenberg model to estimate the Curie temperatures of the bilayer $CrI_3$ under biaxial strains. Although the amplitudes of the Curie temperatures are expected to be overestimated, qualitative investigation of the response of $T_C$ to the biaxial strain is still meaningful.

Figure 7a plots the magnetization (i.e. the average spin moment per $CrI_3$ formula) as a function of temperature for the AB'-stacking bilayer $CrI_3$, from the MC simulations. It can be seen that the average spin moment sharply decreases from 3 $\mu_B$ to 0 at certain temperature for each strain, corresponding to the $T_C$. The Curie temperatures of the strain-free AB- and AB'-stacking bilayers increase to about 130 K and 100 K (i.e. at $\varepsilon = 0$), respectively, which means that both the interlayer FM exchange interaction in the AB-stacking bilayer $CrI_3$ and the interlayer AFM exchange interaction in the AB'-stacking bilayer $CrI_3$ have positive effect on the FM ordering, compared to the monolayer $CrI_3$. With the strain varies from -6% to 6%, the $T_C$ increases from 11 K to 160 K for the AB-stacking bilayer $CrI_3$ and from 9 K to 130 K for the AB'-stacking bilayer $CrI_3$, as shown in Fig. 7b. This trend is similar to that of $J_1$, indicating that the $T_C$ is mainly determined by the intralayer exchange interactions in the bilayer $CrI_3$, similar to the monolayer $CrI_3$. The FM ordering in the AB-stacking bilayer $CrI_3$ is further enhanced by the interlayer exchange interactions, which leads to higher $T_C$ with respect to the AB'-stacking bilayer $CrI_3$ at the same strain. It is worth noting that the stability of the magnetic ordering depends on not only the $T_C$ but also the MAE. Although the MAEs of the bilayer $CrI_3$ under extensile biaxial strain decreases slightly relative to strain-free case (see Fig. 5), the MAEs are already large enough to ensure the magnetic ordering bellow Curie temperatures for both AB- and AB'-stacking bilayers.

Expect for the biaxial strain, the hydrostatic pressure may also modify the magnetism of the bilayer $CrI_3$ [20,21]. Because the interlayer distance is much easier to be changed than the lateral lattice constant, the hydrostatic pressure can be regarded as the compressive strain on the octahedron along the $C_3$ axis as denoted in Fig. 4b. Here, we use the decrement of the distance between the two $CrI_3$ layers ($\Delta d$) to represent the hydrostatic pressure. We firstly calculated the energy difference between the interlayer FM and AFM states, as plotted in Fig. 8a. For the AB-stacking bilayer $CrI_3$, $\Delta E$ keeps positive and increases rapidly as $\Delta d$ increases, which indicates that the interlayer exchange is strongly FM. This can also be manifested by the rapidly increasing $J_2$ and positive $J_3$ as shown in Fig. 8b. For the AB'-stacking bilayer $CrI_3$, $\Delta E$ increases slightly first, then decreases strikingly when $\Delta d > 1.2$ Å. In addition, the sign of $\Delta E$ at $\Delta d = 0.9$ and 1.2 Å is positive, implying that the AB'-stacking bilayer $CrI_3$ favors the interlayer AFM state at this situation.

The intralayer exchange interactions in both AB- and AB'-stacking bilayers remain FM, as manifested by the positive values of $J_1$ in Fig. 8b. Moreover, $J_1$ increases monotonically as $\Delta d$ increases for both systems, and the magnitudes of $J_1$ are much larger than those with the biaxial strain. Therefore, the Curie temperatures increase significantly up to 320 K and 280 K respectively for the AB- and AB'-stacking bilayers, as shown in Fig. 9a. Although $T_C$ is overestimated by the Heisenberg model, the relative enlargement could be instructive. Considering that the amplitudes of $T_C$ of the AB- and AB'-stacking bilayers have been enlarged by about 2.5 times at Δd = 1.8 Å compared to those at pressure-free systems, $T_C$ over 100 K could be expected in real experimental measurement.

The MAE of the AB-stacking bilayer $CrI_3$ decreases monotonically as Δd increases, as shown in Fig. 9b. For the AB'-stacking bilayer $CrI_3$, the MAE decreases first, then increases at Δd = 0.9 Å, and decreases again when Δd > 0.9 Å. In fact, the AB'-stacking bilayer $CrI_3$ favors the interlayer FM state at Δd = 0.9 and 1.2 Å as seen in Fig. 8a, so that the MAEs behave differently from the other points. If the AB'-stacking bilayer $CrI_3$ is in the interlayer AFM state, the MAEs follow the trend formed by the other points, as manifested by the open triangles and connected by the blue dashed lines in Fig. 9b. When Δd > 1.5 Å, the MAEs of both AB- and AB'-stacking bilayers are negative, indicating in-plane MA. Nonetheless, the amplitudes of the MAEs are comparable to or even larger than those of the press-free systems, especially for Δd = 1.8 Å, hence the magnetic ordering could still be retained bellow $T_C$.

## IV. Conclusions

In summary, we investigated the electronic and magnetic properties of two types of bilayer $CrI_3$ under biaxial strain and hydrostatic pressure, based on first-principles calculations and Monto Carlo simulation. We found that the band gaps, MAEs, intralayer and interlayer exchange interactions, and Curie temperatures can be tuned by the biaxial strain and hydrostatic pressure. The band gap decreases (increases) under compressive (extensile) strain, which is related to the symmetry of the $CrI_6$ octahedron. The MAE increases under compressive strain, but decreases under extensile strain and hydrostatic pressure. The interlayer FM coupling of the AB-stacking bilayer $CrI_3$ is robust under both biaxial strain and hydrostatic pressure. The interlayer coupling of the AB'-stacking bilayer $CrI_3$ is sensitive to the biaxial strain. It prefers the interlayer AFM state under extensile strain and small compressive strain (ε > -2%), but turns to the interlayer FM state for larger compressive stain. For both bilayers, the intralayer magnetic ordering may undergo a FM-to-AFM transition under large compressive strain (ε < -6%). The interlayer exchange interaction may be significantly enhanced by the hydrostatic pressure, which may boost the Curie temperature over 100 K.

## Acknowledgements

This work is supported by the National Natural Science Foundation of China (11574223) and the Six Talent Peaks Project of Jiangsu Province (2019-XCL-081).

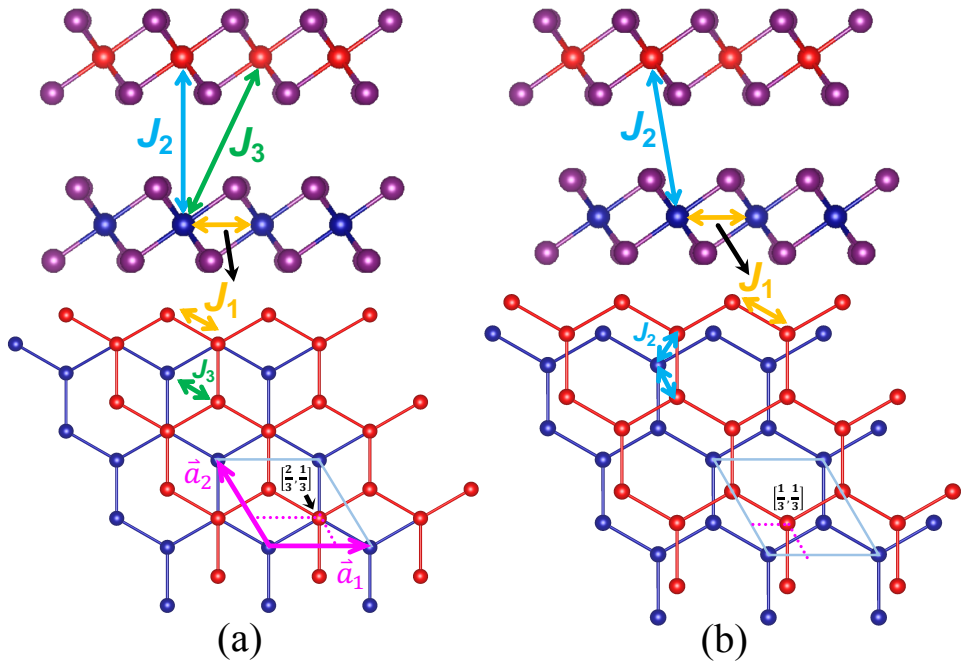

Figure 1. Side and top views of the atomic structure of bilayer CrI$_3$ with (a) AB-stacking order and (b) AB'-stacking order. The red and blue spheres stand for the Cr atoms in the top and bottom layers, respectively. The purple spheres stand for the I atoms. The intralayer ($J_1$) and interlayer ($J_2$ and $J_3$) exchange parameters are indicated. In the bottom layers, only the Cr atoms are shown, and the light blue parallelograms denote the unit cells of the AB- and AB'-stacking bilayers. The purple arrows mark the lattice vectors, and the purple dotted lines indicate the lateral shift (in fractional coordinates) of the top layer (red) with respect to the bottom layer.

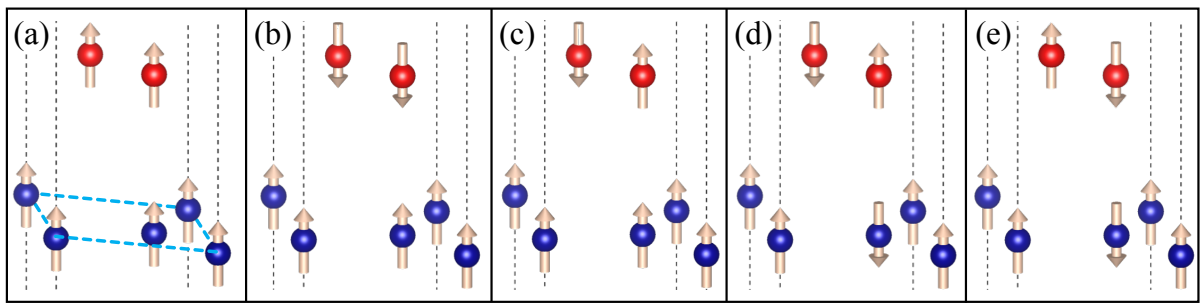

Figure 2. Magnetic configurations of AB-stacking bilayer CrI$_3$. The dashed parallelogram denotes the in-plane unit cell. (a) Intralayer and interlayer ferromagnetism. (b) Intralayer ferromagnetism and interlayer antiferromagnetism. (c) Interlayer ferrimagnetism. The bottom CrI$_3$ layer is ferromagnetic, while the top CrI$_3$ layer is antiferromagnetic. (d) Intralayer and interlayer antiferromagnetism. (e) Intralayer antiferromagnetism and interlayer ferromagnetism. The magnetic configurations of AB'-stacking bilayer CrI$_3$ are similar.

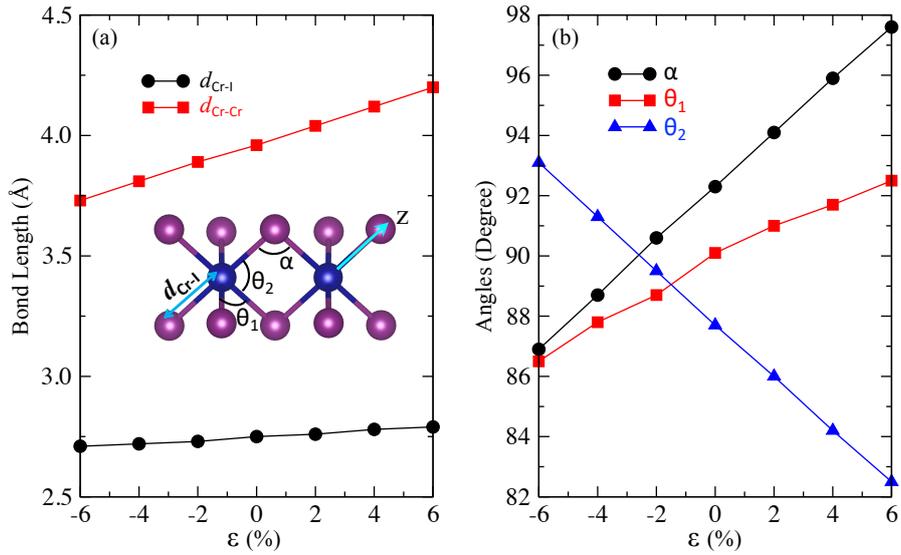

Figure 3. Structure parameters of the AB-stacking bilayer $CrI_3$ under biaxial strain. (a) The Cr-I bond length ($d_{Cr-I}$) and the distance between the nearest Cr atoms ($d_{Cr-Cr}$). (b) Two types of the I-Cr-I angles and Cr-I-Cr angle. The z axis of the local coordinate frame is indicated.

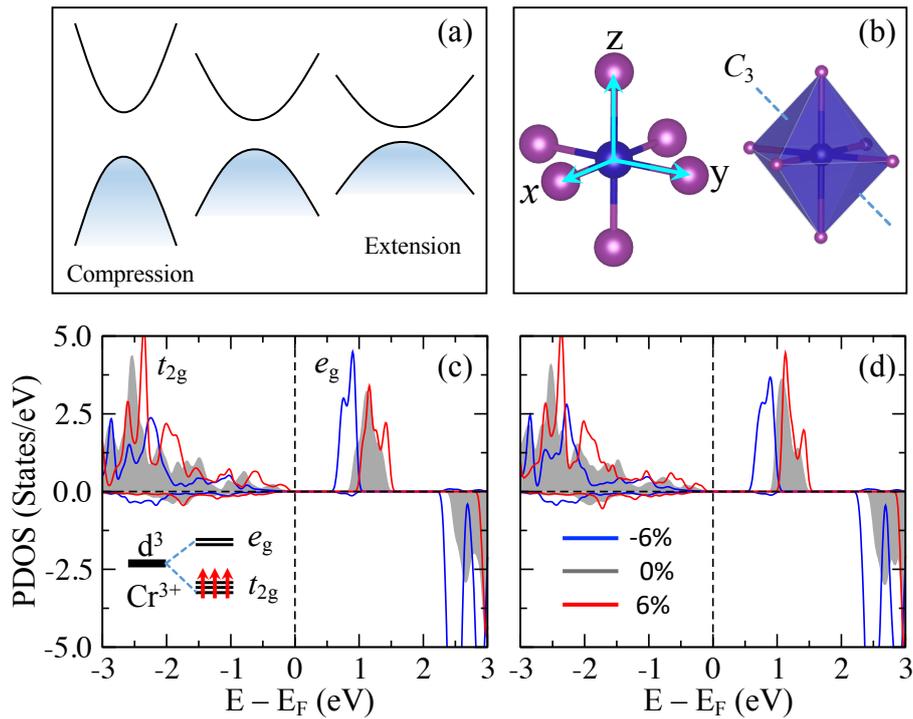

Figure 4. (a) Sketch of the conventional evolution of the band structure under biaxial strain. (b) One $CrI_6$ octahedron with the local coordinate frame. The dashed line indicates the $C_3$ rotational axis and it is the direction of the equivalent strain on the octahedron to the biaxial strain on the unit cell. (c) and (d) Projected density of states (PDOS) of the 3d orbital of the $Cr^{3+}$ ion under biaxial strains for the AB- and AB'-stacking bilayers, respectively. Positive (negative) sign refers to spin majority (minority) channel. The inset in (c) shows the schematic energy splitting of the 3d orbitals of the $Cr^{3+}$ ion.

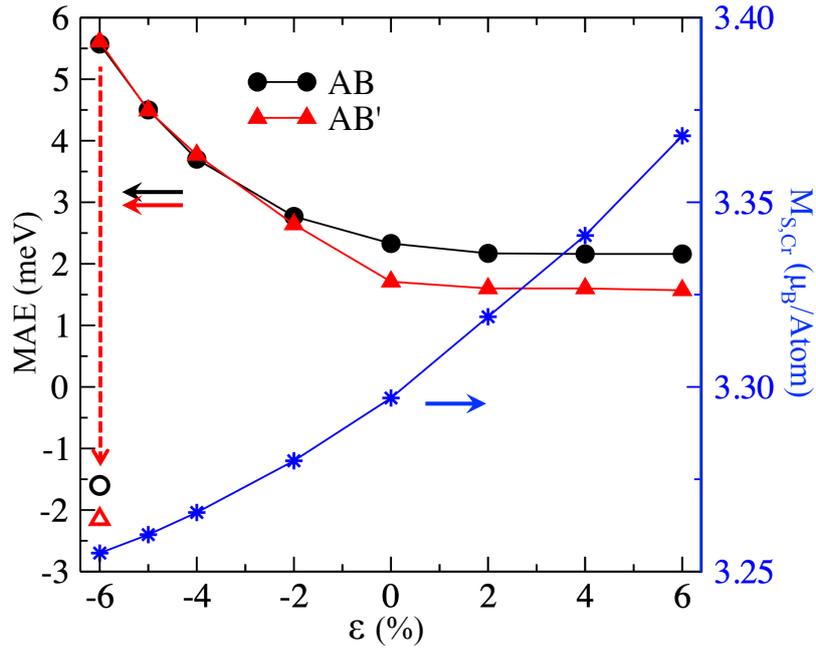

Figure 5. Magnetic anisotropy energy (MAE) and local magnetic spin moment on Cr atom ($M_{S,Cr}$) of the AB-stacking bilayer $CrI_3$ under biaxial strain. The dashed arrow indicates the change of MAE when the FM-to-AFM transition occurs.

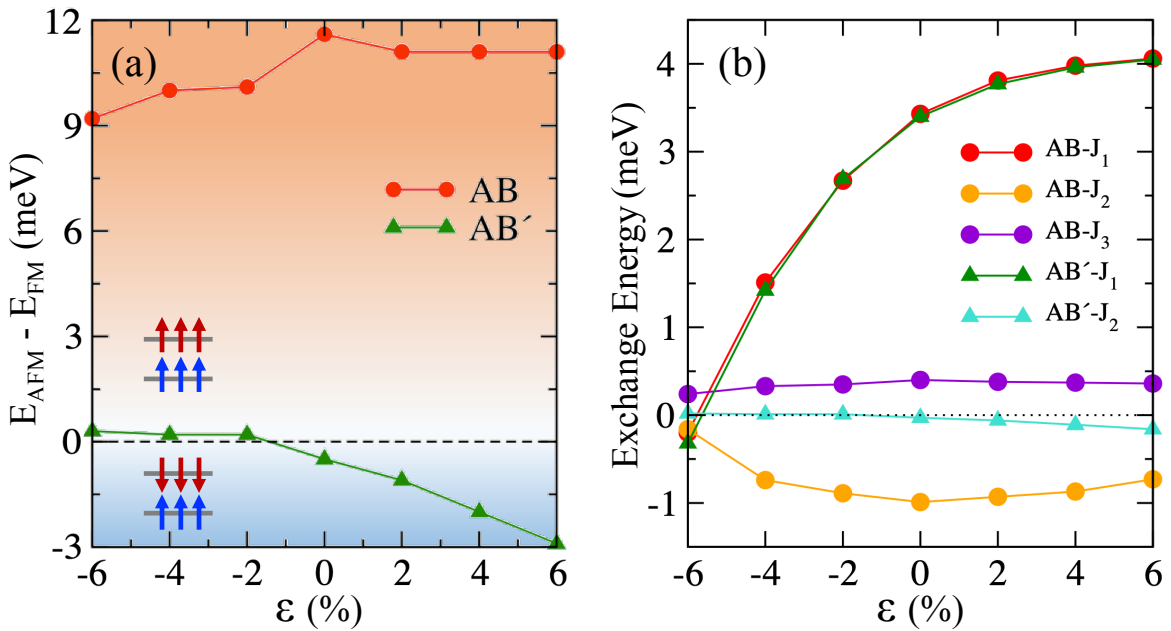

Figure 6. (a) Energy difference between the interlayer FM and AFM states of the bilayer $CrI_3$ under biaxial strain. (b) Exchange interaction energies between the Cr atoms as indicated in Figure 1.

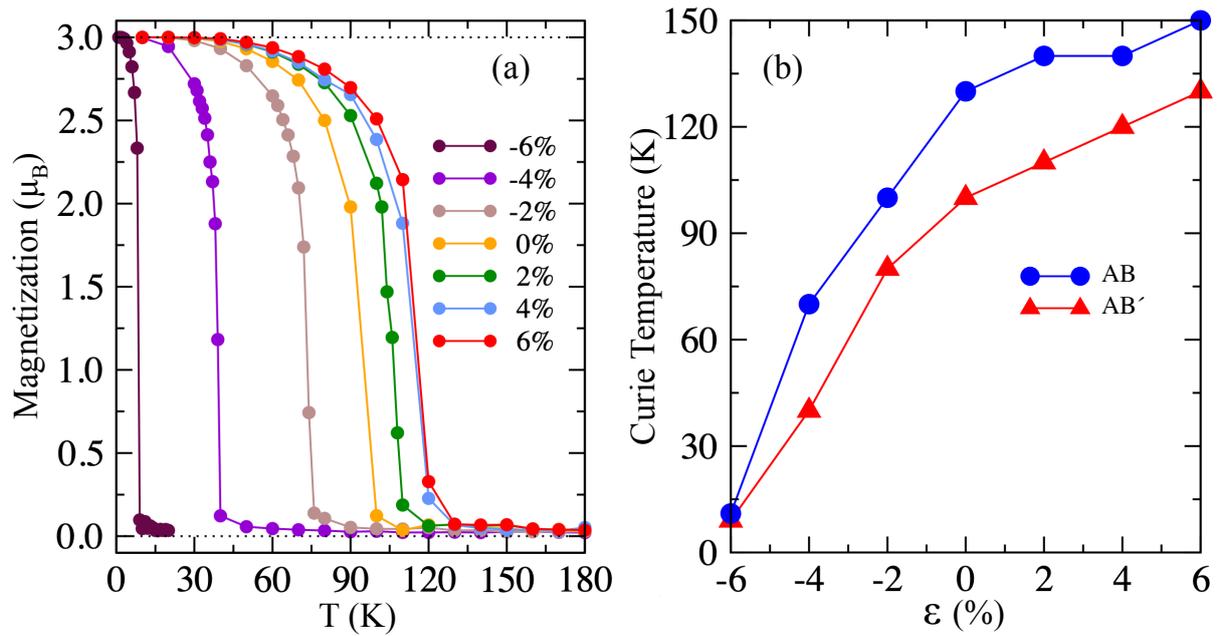

Figure 7. (a) Magnetization (average spin moment per CrI$_3$ formula) of the AB'-stacking bilayer CrI$_3$ at different temperatures. (b) Curie temperature of the bilayer CrI$_3$ under biaxial strain.

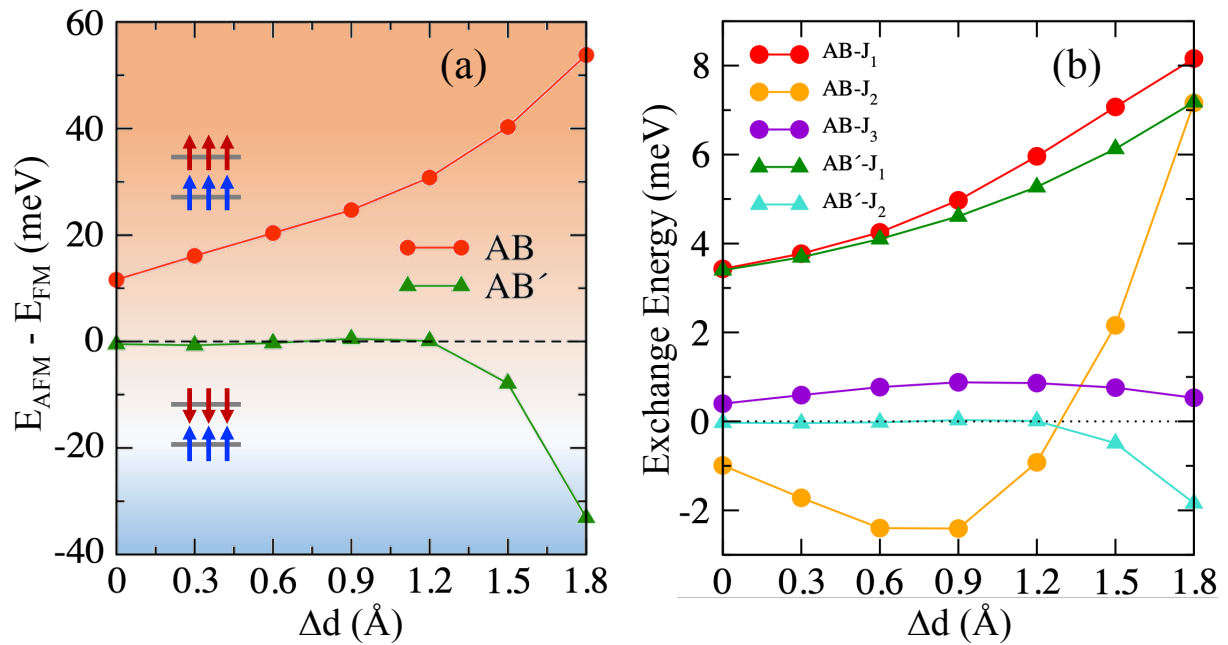

Figure 8. (a) Energy difference between the interlayer FM and AFM states of the bilayer CrI$_3$ under hydrostatic pressure. (b) Exchange interaction energies between the Cr atoms as indicated in Figure 1.

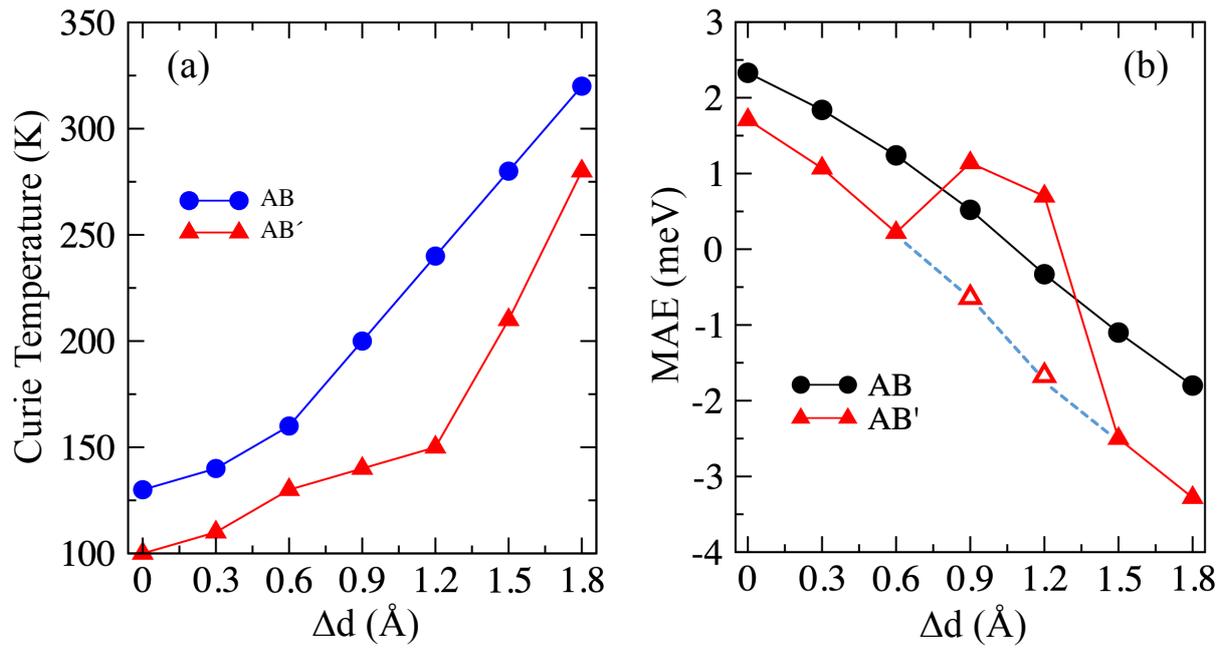

Figure 9. (a) Curie temperature and (b) magnetic anisotropy energy (MAE) of the bilayer CrI$_3$ under hydrostatic pressure. The open triangles were calculated with interlayer AFM state for the AB'-stacking bilayer CrI$_3$.